\documentclass[usenatbib]{mn2e}
\usepackage{graphicx}
\usepackage{natbib}
\usepackage{amssymb}
\usepackage{aas_macros}
\usepackage{lscape}
\usepackage{txfonts}

\title[The extended structure of IC10]{The red extended structure of IC10, the nearest blue compact galaxy}

\author[Gerbrandt et al.]{Stephanie A. N. Gerbrandt$^{1,2}$ Alan W. McConnachie$^2$ \& Mike  
Irwin$^3$\\$^1$ Department of Physics and Astronomy, University of British Columbia, 6224 Agricultural Road, Vancouver, B.C., V6T 1Z1, Canada\\$^2$NRC Herzberg Astronomy and Astrophysics, 5071 West Saanich Road, Victoria, B.C., V9E 2E7, Canada\\${^3}$ Institute of Astronomy,  Madingley  Road,  Cambridge, CB3  0HA,  U.K.\\} 

\begin{document}

\maketitle

\begin{abstract} 

The Local Group starburst galaxy IC10 is the closest example of a blue compact galaxy. Here, we use optical $gi$ imaging from CFHT/MegaCam and near infra-red $JHK$ imaging from UKIRT/WFCAM to conduct a comprehensive survey of the structure of IC10. We examine the spatial distribution of its resolved young, intermediate and old stellar populations to large radius and low effective surface brightness levels. Akin to other dwarfs with multiple populations of different ages, stellar populations of decreasing average age are increasingly concentrated in this galaxy. We find that the young, star-bursting population, and the AGB population, are both offset from the geometric center of the older RGB population by a few hundred parsecs, implying that the younger star formation occurred significantly away from the center of the galaxy. The RGB population traces an extended structure that is typical of blue compact galaxies, with an effective radius of $\sim 5.75$\,arcmins ($\sim1.25$\,kpc). These measurements show that IC10 is much more extended than has previously been realized, and this blue compact galaxy is one of the most extended dwarf galaxies in the Local Group. The outermost isophotes of this galaxy are very regular in shape and essentially circular in morphology. Based on this analysis, we do not find any evidence to suggest that IC10 has undergone a recent, significant, interaction with an unknown companion. 

\end{abstract}


\section{Introduction}

The Local Group is home to a large number of galaxies, the overwhelming number of which are dwarfs. The diversity of star formation histories within this population is well established (e.g., \citealt{mateo1998a, weisz2014}) and the origin of this diversity is the subject of considerable effort that intersects several different astronomical topics. These range from stellar evolution to galaxy formation, including the role of feedback, and the interplay between secular and environmental dependencies on (dwarf) galactic scales. However, of the more than 70 galaxies within a megaparsec or so of the Milky Way, only one is classified as a ``starburst'' system, IC10.

IC10 occupies an awkward position close to the Galactic plane ($b=-3.3^\circ$), that has made study of its optical component notoriously difficult. It has both a high foreground and internal reddening (\citealt{massey1995, sanna2008, kim2009}), but distance estimates have converged to place it in the range of 700 -- 800kpc from the Milky Way (\citealt{sakai1999, demers2004, tully2006, sanna2008, kim2009}). Being in the direction of Cassiopeia, this distance places it as an outlying member of the M31 subgroup. It is one of the most luminous of the Local Group dwarf galaxies ($M_V\simeq -15$) and it has long been known to be vigorously forming stars; its overall star formation rate is the highest in the Local Group. Its high H$\alpha$ luminosity (e.g., \citealt{kennicutt2008}) and an extremely high surface density of Wolf-Rayet stars that is approximately a factor of 20 times larger than the LMC (\citealt{massey1995, massey2002}) lead it to be classified as a starburst galaxy. \citealt{richer2001} argue that it also meets the criteria to be classified as a blue compact dwarf (BCD). Whatever its classification, IC10's star formation rate in comparison to our other nearby neighbours is unusually high.

\cite{shostak1989} showed that IC10 contains significant amounts of gas in a complex structure. At least two components are evidenced; the main rotating disk-like component and an outer, extended envelope that rotates in the opposite sense. \cite{shostak1989} note that this galaxy ``is obviously in dynamical disarray'', and consider possible origins, including a merger with a companion of roughly equal HI mass. However, these authors also point out that such a morphology is not unique; they argue that Sextans A and NGC4449 both show similar structures with extended HI envelopes surrounding central rotating disks, and suggest that this is a sign of previously heated material being re-accreted into the center of the potential.  It is notable that, since this work, NGC4449 has been shown to be undergoing a merger with a lower mass galaxy (\citealt{martinezdelgado2012}; earlier work by \citealt{hunter1998} suggests this could be the second such merger for NGC4449). The complex gaseous morphology of IC10 prompted \cite{wilcots1998} to suggest that it is a galaxy still forming via the accretion of gas.

The suggestion that IC10 is undergoing an interaction is an attractive solution to simultaneously explain its gas kinematics and its current intense star formation activity. Until recently, it was unclear if such a process was plausible for IC10. However, IC10 is one of the few galaxies in the Local Group with a measured proper motion through measurements of masers using the Very Long Baseline Array (\citealt{brunthaler2007}). New orbital solutions suggest that its last M31 pericentric passage was nearly 2 Gyrs ago (\citealt{nidever2013}; the galaxy is now $\sim 250$\,kpc away from M31, \citealt{mcconnachie2012}). This does not match well with the estimated age of the starburst (estimated to be of order $\sim 10$\,Myrs by \citealt{massey2007}, although the exact age is uncertain and it could plausibly be older). Its closest known neighbor is possibly the newly discovered Cassiopeia III/Andromeda XXXII dwarf spheroidal (\citealt{martin2013a}), at a distance of $\sim110$\,kpc  and a luminosity of $\sim -12.3$. Recent interactions with known Local Group galaxies therefore seem highly unlikely. Interactions of IC10 with an unknown - or now-destroyed - companion, is still plausible, particularly since IC10 lies in a region of sky where there can be no claims on the completeness of the sample of Local Group galaxies (e.g., \citealt{irwin1994, whiting2007}).

Recently, \cite{nidever2013} have analysed Green Bank Telescope (GBT) HI surveys in the region of IC10. They present evidence of an extended feature stretching 1.3\,degrees to the northwest of IC10, that displays a large velocity gradient and that ends in a high column-density clump with a coherent velocity. This feature seems to lead IC10 in its orbit  (i.e., it is not likely due to ram-pressure stripping), and \cite{nidever2013} argue that the most plausible explanation is that the HI is associated with an interaction or merger with another, unknown, dwarf galaxy, that presumably has a relatively low surface brightness.

Such an interaction as proposed by \cite{nidever2013} has the potential to leave imprints not only on the gas distribution, but also on the stellar distribution at large radius, since it is feasible the interaction may have produced a tidal stream, or perturbed the outer regions of the galaxy. Indeed, \cite{loose1986} showed that blue compact galaxies are often embedded within a large extended, low surface brightness stellar distribution. \cite{sanna2010} showed  that IC10 is similarly embedded within an extended distribution of red giant stars, that extends to at least 20 arcmins. However, the two dimensional structure and morphology of this component is not yet measured. In light of recent claims, it is worth revisiting the extended structure of this intriguing galaxy.

The structure of this paper is as follows: Section~2 provides a description of the CFHT/MegaCam and UKIRT/WFCAM datasets that provide the basis for our analysis, and provides a broad overview of the features present in the associated color-magnitude diagrams. Section 3 details our attempts to deal with foreground contamination and to estimate a distance to IC10 using the near-IR dataset. Section~4 examines the spatial distribution of the young, intermediate and old stellar populations, with particular emphasis on the low surface brightness structure of the galaxy. Section~5 discusses our findings, and Section 6 provides a summary of our results.

\section{Data acquisition, processing and characterization}

The NIR dataset was obtained using the Wide Field Camera on the 3.8m United Kingdom Infrared Telescope during a single observing run in October 2008. These data were taken as part of program to obtain deep NIR imaging for every Local Group galaxy visible from Maunakea (P.I. Irwin; previous publications from this survey include \citealt{cioni2008} and \citealt{sibbons2012}).  A detailed description of WFCAM can be found in \cite{casali2001}. Briefly, WFCAM comprises four non-contiguous Rockwell-Hawaii-II infrared detector arrays (HgCdTe $2048\times2048$\, pixels) that can be used to observe a contiguous ``tile'' of 0.75deg$^2$. For IC10, the exposure time of each tile in $JHK$ was 200s, made up of 5 different dither positions with 2x2 microstepping at each position to interleave a 0.2'' sampled image. Each individual exposure was 10s. Four of these pointings were used to make a contiguous tile. The seeing was generally very good, with an average IQ of [$0.74", 0.65", 0.70"$] for $JHK$.

The WFCAM pipeline at the Institute of Astronomy in Cambridge was used to process the data, and follows the same procedures as described in \cite{sibbons2012}. As in that paper, the magnitudes quoted for each source are based on aperture photometry, with zero-points calibrated against 2MASS although they are not transformed into the 2MASS system (\citealt{hodgkin2009}). Rather, they remain on the WFCAM instrumental (Vega) system, with the relevant transformation equations given in \cite{hodgkin2009}.

The optical dataset is archival $gi$ (AB mags) data from the MegaCam wide field camera on the Canada-France-Hawaii Telescope, that has previously been presented in \cite{sanna2010}. MegaCam is a mosaic of 36 (usable) individual $2048\times4612$ CCDs  arranged in a $9\times4$ grid, resulting in a contiguous field of view of $0.96\times0.94$\,degrees with a pixel scale of 0.187''/pixel. The exposures for IC10  are the result of the co-addition of three dithered sub-exposures ($3\times700$s in $g$, $3\times400$s in $i$).  CFHT/MegaCam data are pre-processed by CFHT staff using the Elixir pipeline, which accomplishes the usual bias, flat, and fringe corrections and also determines the photometric zero point of the observations. These images were then processed using a version of the Cambridge photometry pipeline (\citealt{irwin2001}) adapted for CFHT/MegaCam observations, and which has been used extensively for the Pan-Andromeda Archaeological Survey (PAndAS; \citealt{mcconnachie2009b}). The pipeline includes re-registration, stacking, catalog generation, and object morphological classification, and creates band-merged $gi$ products for use in the subsequent analysis. The image quality of the $i-$band data is consistently better than the $g-$band data ($\sim0.8-0.9$'' instead of $\sim1.0-1.1$''). We found that a deeper and more consistent stellar catalog was produced when we used objects identified in the $i-$band to force photometry in the $g-$band. The object classification fo the CFHT data is therefore taken from the $i-$band only.

The final WFCAM merged dataset contains a total of 224\,679 sources, of which 130\,581 are confidently classified as stellar (within 2 sigma of the stellar locus) in at least 2 of the three bands.  The final MegaCam catalog contains a total of 290\,720 sources, of which 210\,156 are confidently classified as stellar (within 2 sigma of the stellar locus) in the $i-$band. 

\begin{figure*}
  \begin{center}
    \includegraphics[angle=270, width=7.5cm]{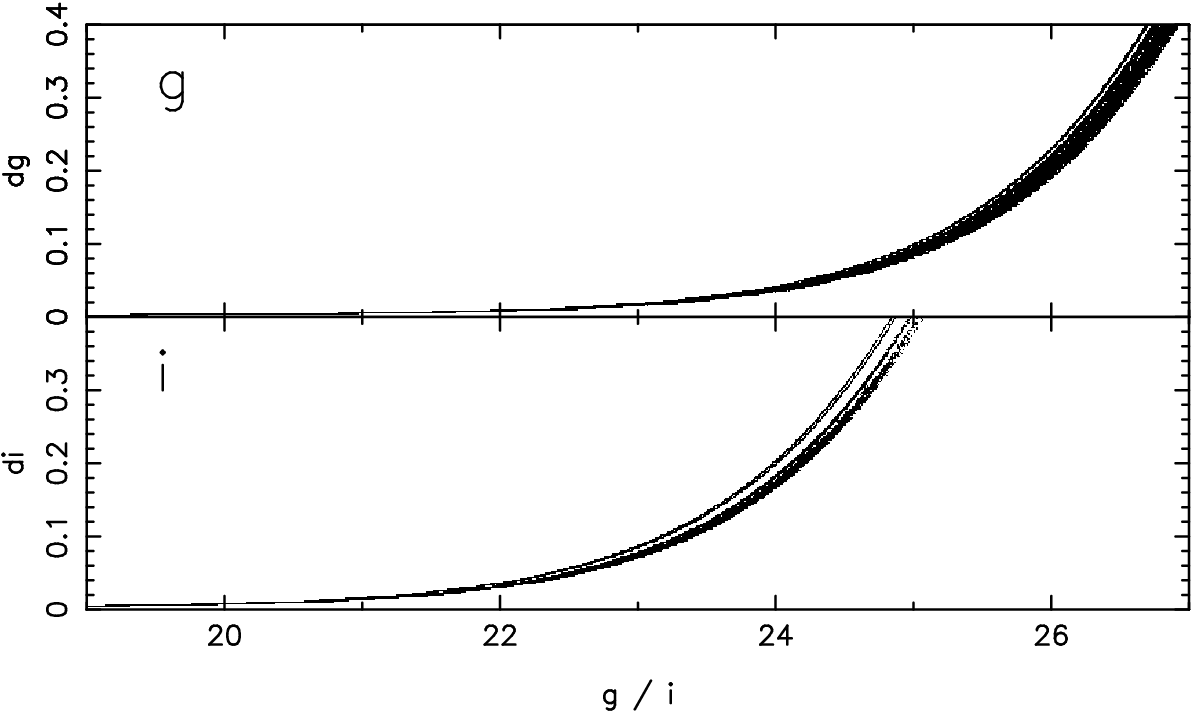}
    \includegraphics[angle=270, width=7.5cm]{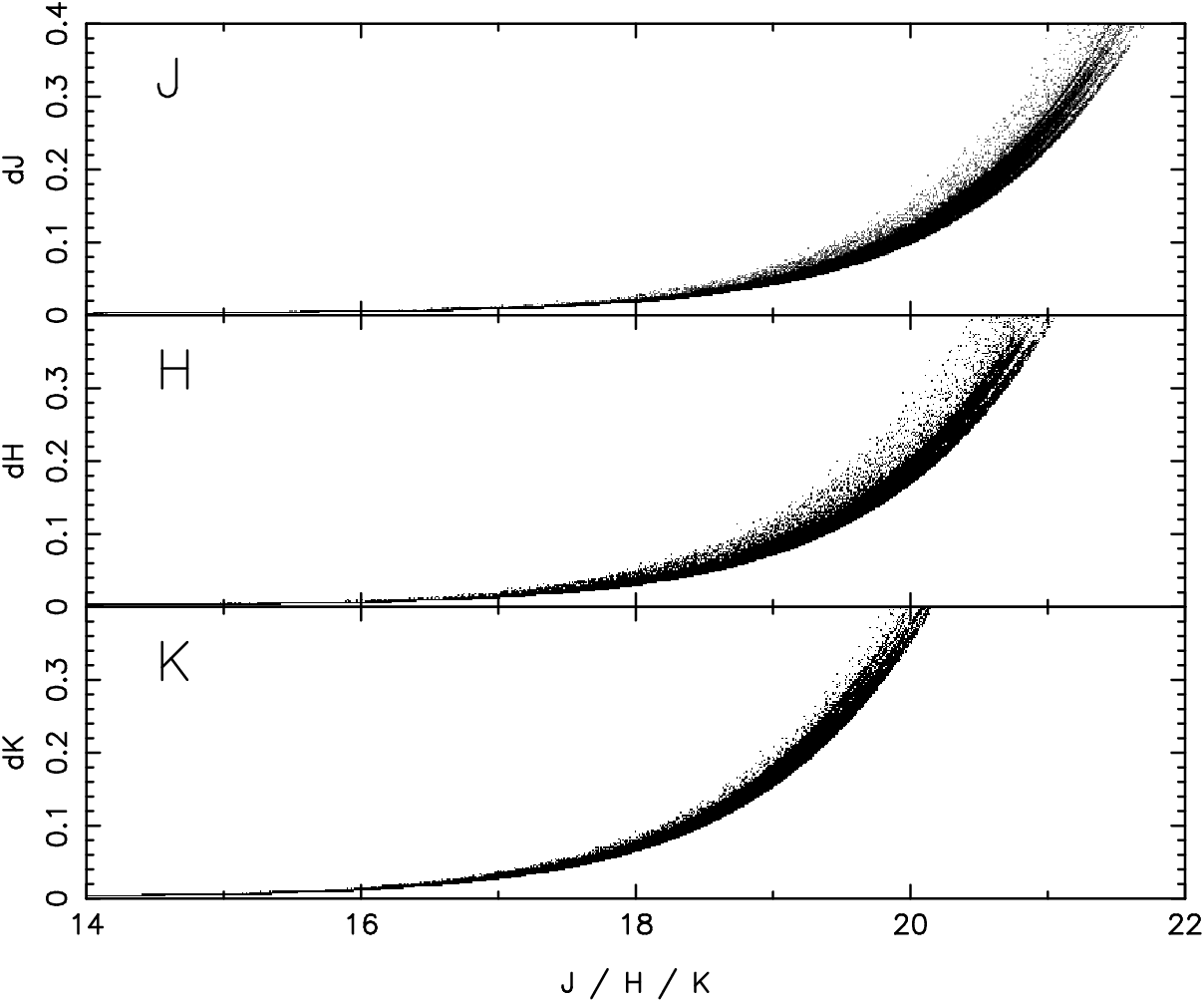}
   \caption {Photometric uncertainties as a function of magnitude for the two optical bands from the CFHT/MegaCam data (left panels) and the three NIR bands from the UKIRT/WFCAM data (right panels).} 
  \end{center}
\end{figure*}

\begin{figure*}
  \begin{center}
    \includegraphics[angle=270, width=7.5cm]{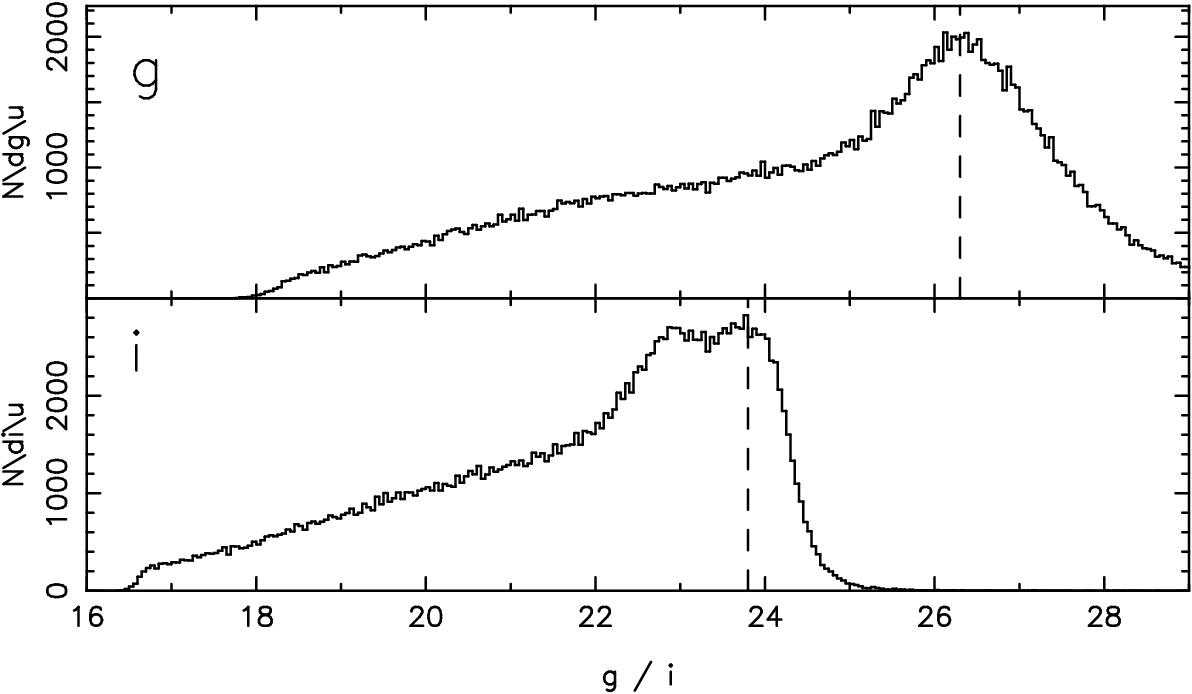}
    \includegraphics[angle=270, width=7.5cm]{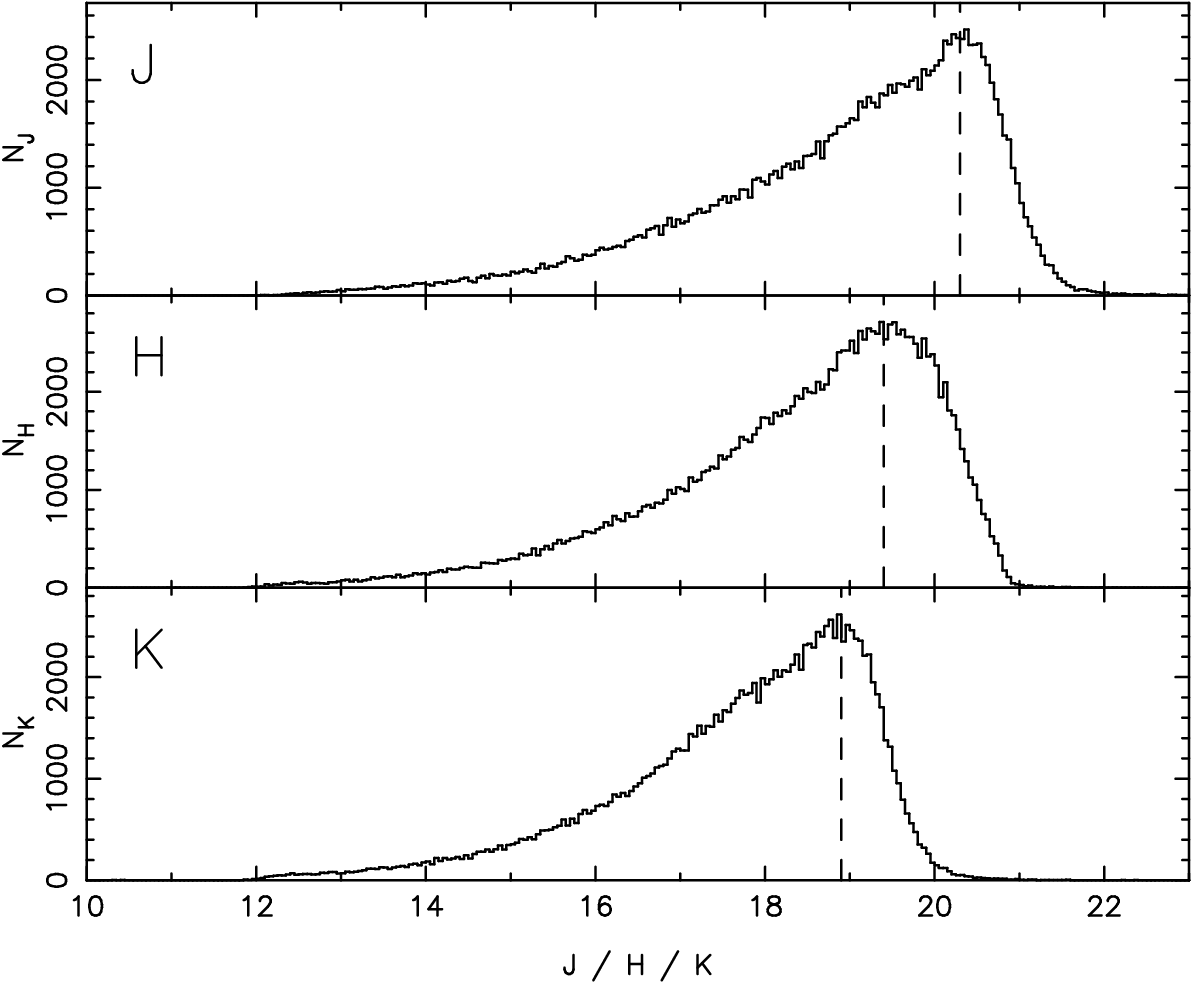}
    \caption {Luminosity functions of all stellar sources identified in the two optical bands from the CFHT/MegaCam data (left panels) and the three NIR bands from the UKIRT/WFCAM data (right panels). For the NIR luminosity functions, stellar classification has been conducted on each band independent of all other bands. For the optical bands, object identification and classification was conducted exclusively using the $i-$band. Vertical dashed lines indicate the turn-over in the luminosity functions, indicating significant incompleteness at these magnitudes.}
  \end{center}
\end{figure*}

Figure~1 shows the derived photometric errors for stars in each of five bands as a function of magnitude. Individual sequences corresponding to individual detectors are particularly noticeable for the MegaCam $i$ bands. In general, our photometric errors are better than 10\% for [$g, i, J, H, K$] $\simeq$ [$25.0, 23.2, 19.8, 19.2, 18.5$]. Figure~2 shows the luminosity functions for stellar objects in each of the 5 bands to illustrate the general depth of the data in each band. Vertical dashed lines at [$g, i, J, H, K$] $\simeq$ [$26.3, 23.8, 20.3, 19.4, 18.9$] indicate the turn-over in the luminosity functions where the incompleteness has become sufficiently large to dominate over the increasing star counts at faint magnitudes. The analyses presented in this paper do not require precise knowledge of the shape of the incompleteness function as a function of magnitude.

Finally, we correct the observed magnitudes for reddening. Normally, we would do this on a star-by-star basis by cross-referencing to the E(B-V) maps derived from IRAS/DIRBE data by \cite{schlegel1998}. However, the low latitude - and consequently extremely high expected foreground reddening - of IC10, means that we are working in a regime in which the reddening maps of \cite{schlegel1998} are known to be inaccurate. For reference, the mean reddening measured for stars in our IC10 fields from \cite{schlegel1998} is E(B-V)$\sim0.65$\,mags (after applying the correction from \cite{bonifacio2000} that is recommended in regions of moderately high reddening). However, independent estimates of the total reddening (i.e., internal plus foreground) in IC10 by other authors suggest the value is nearer 0.7 -- 0.8 (e.g. $0.77\pm 0.07$, \citealt{richer2001}; $0.78\pm0.06$,  \citealt{sanna2008}; but also $0.98\pm0.06$, \citealt{kim2009}). Given these uncertainties, adopting a spatially varying value on a star by star basis is not warranted, and we adopt a constant reddening value of E(B-V)$=0.77\pm 0.07$ (\citealt{richer2001}). In what follows, adoption of different reddening assumptions do not have any qualitative changes on our results.

\section{Color-magnitude diagrams and identification of the stellar populations}

\begin{figure*}
  \begin{center}
    \includegraphics[angle=270, width=7.5cm]{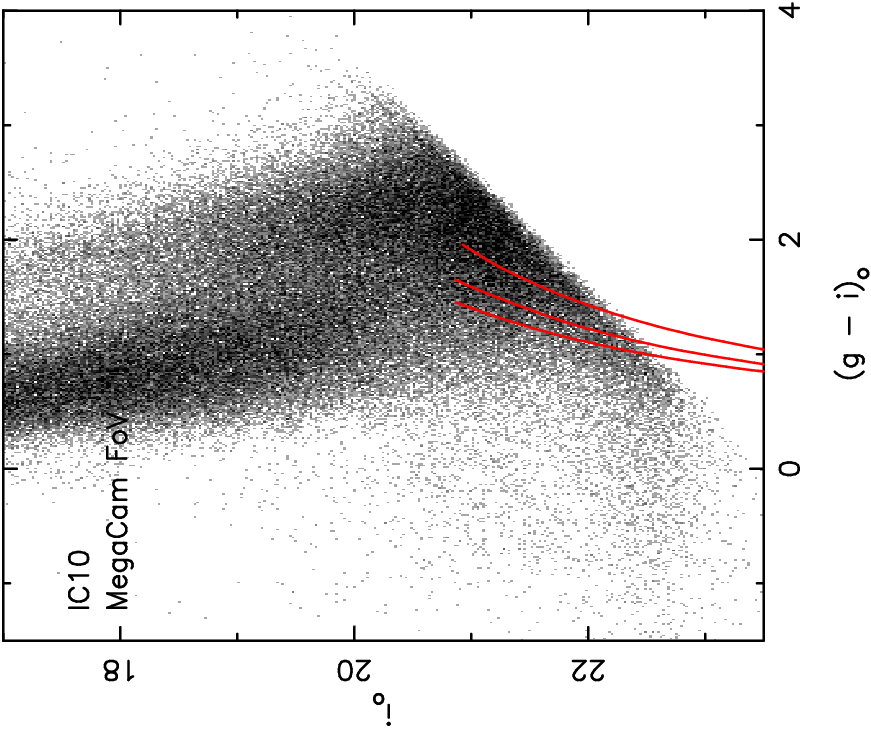}
    \includegraphics[angle=270, width=7.5cm]{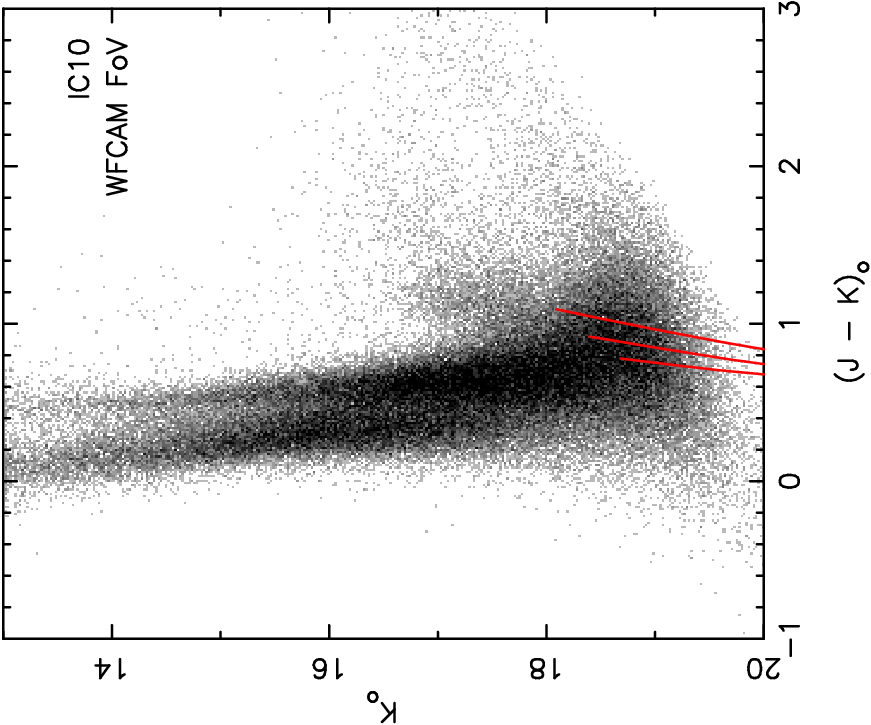}
   \caption {Color-magnitude diagrams (CMDs) of stellar sources in the entire CFHT/MegaCam field of view (left panel; $di\le0.5$, $dg\le0.3$\,mag) and the UKIRT/WFCAM field of view (right panel). These are plotted as Hess diagrams with square-root scaling to better show the structure in these diagrams. Stellar isochrones  corresponding to a 13 Gyr population with [Fe/H] $= -2.0, -1.5$, and -1.0\,dex from the Dartmouth stellar evolution library (\citealt{dotter2008}) are overlaid on both CMDs, shifted to the approximate distance modulus of IC10. The overwhelming fraction of stars in both CMDs are foreground stars in the Galaxy, that swamp the IC10 population.} 
  \end{center}
\end{figure*}

\begin{figure*}
  \begin{center}
    \includegraphics[angle=270, width=7.5cm]{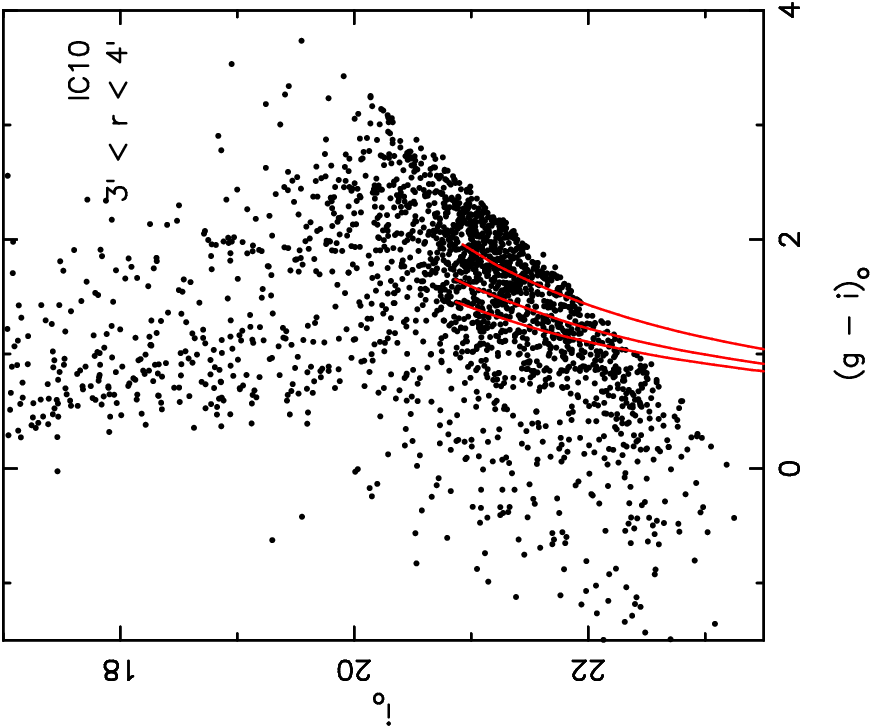}
    \includegraphics[angle=270, width=7.5cm]{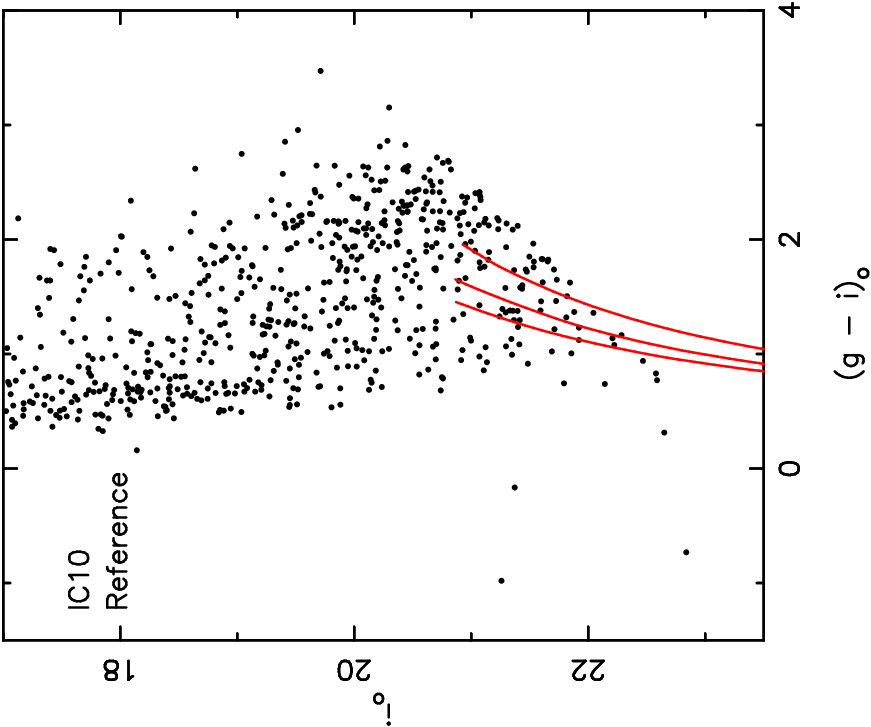}
   \caption {The left panel shows the optical color-magnitude diagram (CMD) from the CFHT/MegaCam for stars located in a circular annulus between 2 and 3 arcmins from the center of IC10. This annulus has been chosen to sample an area where the IC10 population is strong in comparison to the Galactic foreground, while avoiding the very central regions where crowding is significant. The right panel is a reference field, at a random location in the field, offset from IC10 and with the same total area. It shows the presence of the Galactic foreground, but an absence of any other stellar populations. Both CMDs show only stellar sources with $di\le0.5$, $dg\le0.3$\,mag} 
  \end{center}
\end{figure*}

\begin{figure*}
  \begin{center}
    \includegraphics[angle=270, width=12cm]{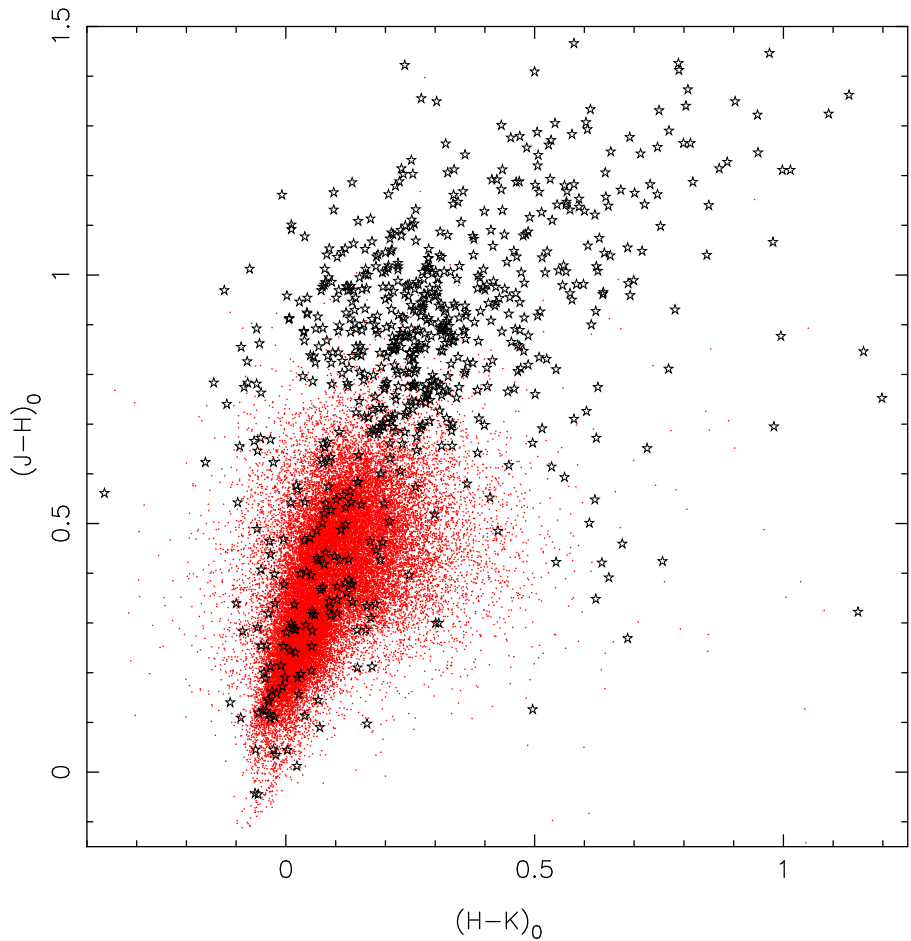}
    \caption {Near-IR color-color diagram of stellar sources in the UKIRT/WFCAM field with uncertainties less than 0.1mags in each filter. Red points shows the location of stars located more than 20 arcmins from the center of IC10, and will predominantly consist of Galactic foreground dwarfs Black points  show the positions of stars within 2 arcmins of the center of IC10, and will consist predominantly of giants in IC10.  The overwhelming majority of the candidate dwarfs lie blue-ward of $(J-H)_0 = 0.7$, whereas only a small fraction of the candidate giants are blue-ward of this limit.} 
  \end{center}
\end{figure*}

\begin{figure*}
  \begin{center}
    \includegraphics[angle=270, width=7.5cm]{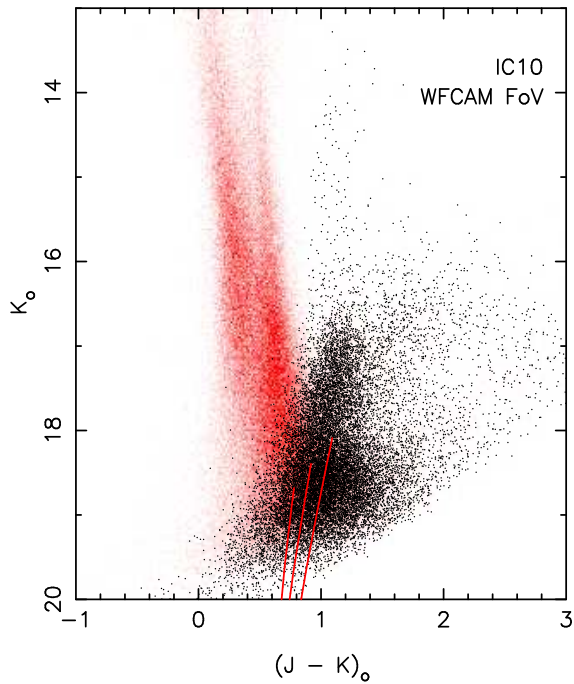}
   \caption {NIR CMD of stellar sources in the UKIRT/WFCAM. The red density plot shows all sources with $(J-H)_0 < 0.7$, and which is dominated by the Galactic foreground population. Black points are stars with  $(J-H)_0 \ge 0.7$, and are mostly genuine giant stars in IC10. Stellar isochrones  corresponding to a 13 Gyr population with [Fe/H] $= -2.0, -1.5$, and -1.0\,dex from the Dartmouth stellar evolution library (\citealt{dotter2008}) are overlaid. Stars brighter than the RGB are mostly asymptotic giant branch stars (AGB) stars. The cloud of stars redward of $(J-K)_0\sim 1.4$ between $16<K_0<18$ are C-rich AGB stars, and O-rich AGB stars are visible in the bright vertical sequence at $(J-K)_0\sim 1$} 
  \end{center}
\end{figure*}

Figure~3 shows the extinction-corrected color magnitude diagrams of stellar sources identified in the full CFHT/MegaCam (left panel; $di\le0.5$, $dg\le0.3$\,mags)  and UKIRT/WFCAM (right panel) fields. Stellar isochrones  corresponding to a 13 Gyr population with [Fe/H] $= -2.0, -1.5$, and -1.0\,dex from the Dartmouth stellar evolution library (\citealt{dotter2008}) are overlaid on both CMDs, shifted to a distance of 750\,kpc, to highlight the approximate location of red giant branch stars in these plots. Both panels of Figure~3 are dominated not by stars in IC10, but by the Galactic foreground that lies along our line of sight to this very low-latitude galaxy ($b=-3^\circ$). This foreground is visible in the NIR CMD as two ``fingers'' extending over the full magnitude range, centered around $(J-K)_0\simeq 0.2$ and $(J-K)_0\simeq0.6$. The bluer of these features consists predominantly of Galactic main sequence turn-off stars (F-K dwarfs) at a range of distances along our sight-line in the disk and halo. The redder feature consists mostly of disk K~dwarfs, with some disk K~giants (e.g. \citealt{nikolaev2000}). The same features are also present in the optical CMD, where the turn-off stars are the vertical sequence at $(g-i)_0 \simeq 1.0$, and the disk dwarfs form the red cloud of points around $(g-i)_0 \sim 2 - 3$. In both CMDs, there is a notable color gradient in these features, with the fainter objects appearing redder than the brighter objects. This is a distance effect, since the Galactic sources further from us have a higher integrated reddening along our sight-line than the closer sources.

To better increase the contrast between stellar populations intrinsic to IC10 and the Galactic foreground, the left panel of Figure~4 shows the optical CMD for stars within 4 arcmins of the center of IC10 (with $di\le0.5$, $dg\le0.3$\,mags). Detections within the inner 3 arcmins have been removed from this CMD, since detections in this region are highly unreliable due to severe crowding issues. The right panel of Figure~4 shows the optical CMD of stars in a random region of the MegaCam field that samples the same total area as the left panel and is well away from the expected location of IC10. While the foreground populations that so dominate in Figure~3 are still present in both panels, the field centered on IC10 has a clear excess of stars compared to the reference field. In particular, a large number of stars with $(g-i)_0\lesssim 0.5$ are present, and correspond to young main sequence and blue supergiant stars in this galaxy. Further, there is a very large number of points, with colors typically redder than $(g-i)_0 \gtrsim 0.8$ and $i_0 \gtrsim 20$. These are mostly red giant branch stars, as well as younger, more luminous AGB stars. It is also clear, however, that these stars do not occupy a clear, concentrated locus that follows the isochrones, but rather the points cluster in a ``cloud'' around the approximate location of the isochrones. This is due to the uncorrected effects of extinction discussed earlier. For this reason, we consider it wholly unreliable to draw any significant conclusions from the colors or luminosities of these optical data (e.g., such as inferring the TRGB or the metallicity via the color of the RGB).

Considering now the near-IR data, we note that it is possible to remove the overwhelming majority of foreground (dwarf) stars using a near-IR color-cut, as demonstrated in \cite{gullieuszik2008} and \cite{sibbons2012}.  Figure~5 shows a $(J-H)_0, (H-K)_0$ color-color diagram of all stellar sources in the UKIRT/WFCAM field. Only sources with uncertainties less than 0.1mags in each filter are plotted. Red points show the location of all stars located more than 20 arcmins from the center of IC10, and will predominantly consist of Galactic foreground dwarfs. Black points  show the positions of all stars within 2 arcmins of the center of IC10, and will consist predominantly of giants in IC10.  The overwhelming majority of the candidate dwarfs lie blue-ward of $(J-H)_0 = 0.7$, whereas only a small fraction of the candidate giants are blue-ward of this limit. We can adopt $(J-H)_0 = 0.7$ as a color-cut to remove the majority of the foreground population (specifically, 96\% of all foreground stars lie blue-ward of this limit in Figure~4). The removal of the small number of genuine IC10 stars by this cut makes no difference to the subsequent analysis.

Figure~6 shows the same near-IR CMD as in the right panel of Figure~3. The black points indicate all stellar sources with $(J-H)_0 > 0.7$ and are mostly members of IC10. The red density plot shows the location of all stars with  $(J-H)_0 < 0.7$, and are mostly foreground stars, as discussed previously. This CMD now clearly shows distinct populations of AGB stars. In particular, O-type AGB stars are visible in the broad, vertical sequence at around $(J-K)_0 = 1$, and C-type AGB stars are visible at around $16<K_0<18$ in a sequence that extends red-wards in color to $(J-K)_0\sim2$. 

\begin{figure*}
  \begin{center}
    \includegraphics[angle=270, width=12cm]{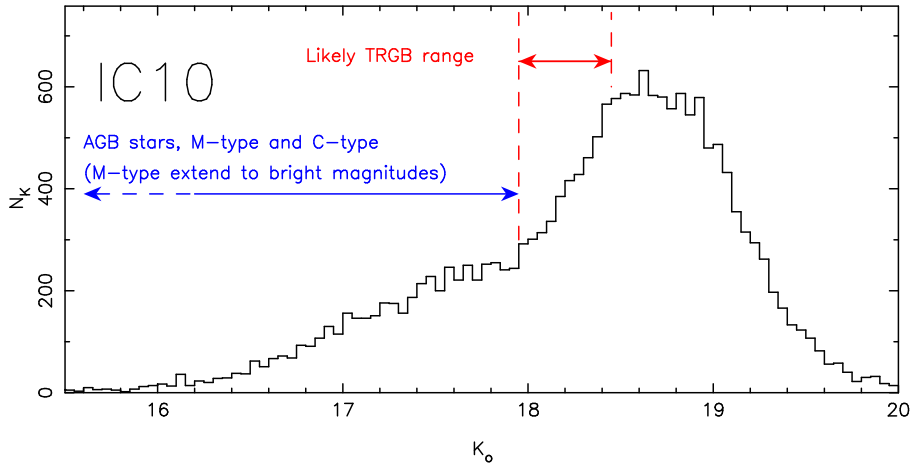}
   \caption {$K_0$ band luminosity function of all stellar sources shown in Figure~6 with $(J-H)_0 \ge 0.7$. The large number of AGB stars in IC10 makes determination of the onset of the RGB very difficult. The likely TRGB range is indicated; the bright-end limit to this range is marked by a change in slope of the luminosity function, and the faint-end limit is set by a large jump in star counts.} 
  \end{center}
\end{figure*}

In addition to the AGB population, RGB stars are also present in Figure~6 and occupy broadly similar colors as the (O-type) AGB stars, but at fainter magnitudes. Figure~7 shows the $K_0$-band luminosity function for the IC10 stars in Figure~6. \cite{bellazzini2004a} provide a calibration for the tip of the red giant branch as a distance indicator using near-IR wavelengths. At these wavelengths, it is a much stronger function of metallicity than in the $I-$band, which complicates its usage for complex stellar populations where stars occupy a broader metallicity range. Nevertheless, identification of plausible values of the tip in the $K$ band provides us with a rough estimate of its distance.

Unlike equivalent luminosity functions for dwarf galaxies that do not have a significant AGB population, the discontinuity in counts due to the tip of the RGB is not clearly defined. In this specific situation, the application of edge-detection techniques (e.g., the Sobel edge detection algorithm, \citealt{lee1993a, sakai1996}) or bayesian or maximum likelihood techniques (e.g., \citealt{makarov2006,conn2011}), that do not explicitly account for the (difficult to model) AGB population, do not help identify any particular feature as due to the tip. By joint inspection of the luminosity function and CMD, however, we can identify a plausible range of values. The largest discontinuity in the luminosity function is at $K\simeq 18.4$, and marks the faintest plausible location we identify as the tip; at brighter magnitudes, there is a steady, roughly constant rise in star counts from $K_0\simeq 16.0$ to $K_0\simeq 17.95$ (that corresponds primarily to the bright AGB populations in the CMD) and then a change in slope to a steeper rise in counts with magnitudes. This change in slope could plausibly be due to the onset of the RGB.

In the 2MASS filter system, \cite{bellazzini2004a} derive the following relationship between TRGB magnitude and overall metallicity for $K$:

\begin{equation}
M_K^{TRGB} = -0.64 [M/H] - 6.93
\end{equation}

\noindent We adopt a metallicity value of [Fe/H] $=$ [M/H] $= -1.3$\,dex for IC10 (\citealt{tikhonov2009}), which yields an absolute K-band magnitude of -6.1\,mags (the correction between the WFCAM and 2MASS $K-$band systems is negligible). The metallicity estimated by \cite{tikhonov2009} is based on the RGB and so is appropriate for our estimate, although we also note that higher metallicity values exist based on planetary nebulae and HII regions, that suggest its metallicity is closer to that of the SMC (e.g., \citealt{richer2001, magrini2009}). A change in the metallicity of $\sim0.2$\,dex will change the absolute magnitude of the TRGB by $\sim0.1$\,mags. For the suspected TRGB location of $17.95 \lesssim K \lesssim 18.45$, this corresponds to a distance modulus of $24.05 \lesssim (m-M)_0 \lesssim 24.55$ ($650 \lesssim D \lesssim 810$\,kpc). This is consistent with previous estimates (e.g.,  660\,kpc, \citealt{sakai1999}; 741\,kpc, \citealt{demers2004}; 795\,kpc, \citealt{tully2006}; 798\,kpc, \citealt{sanna2008}; 715\,kpc, \citealt{kim2009}).

\section{Structural properties}

\subsection{Stellar density maps and shapes}

\begin{figure*}
  \begin{center}
    \includegraphics[angle=270, width=8.5cm]{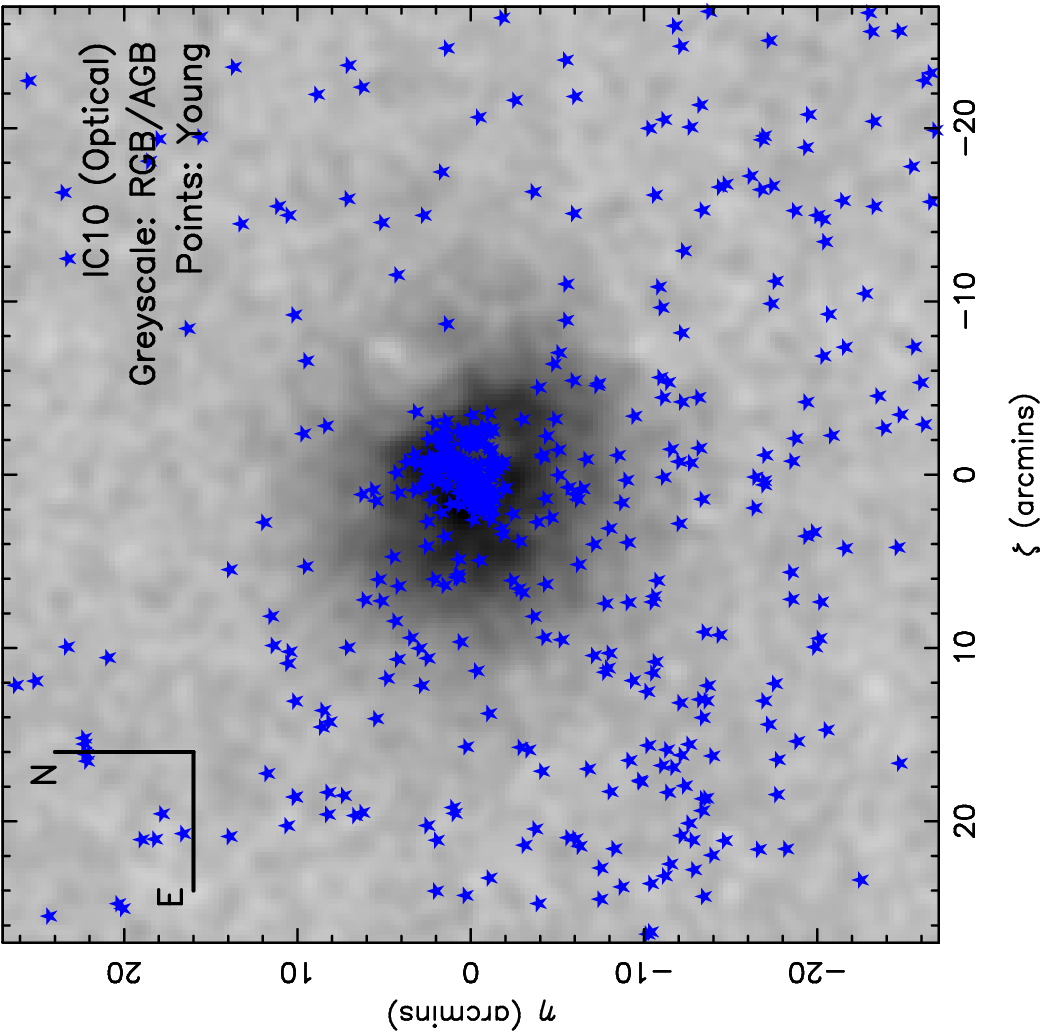}
    \includegraphics[angle=270, width=8.5cm]{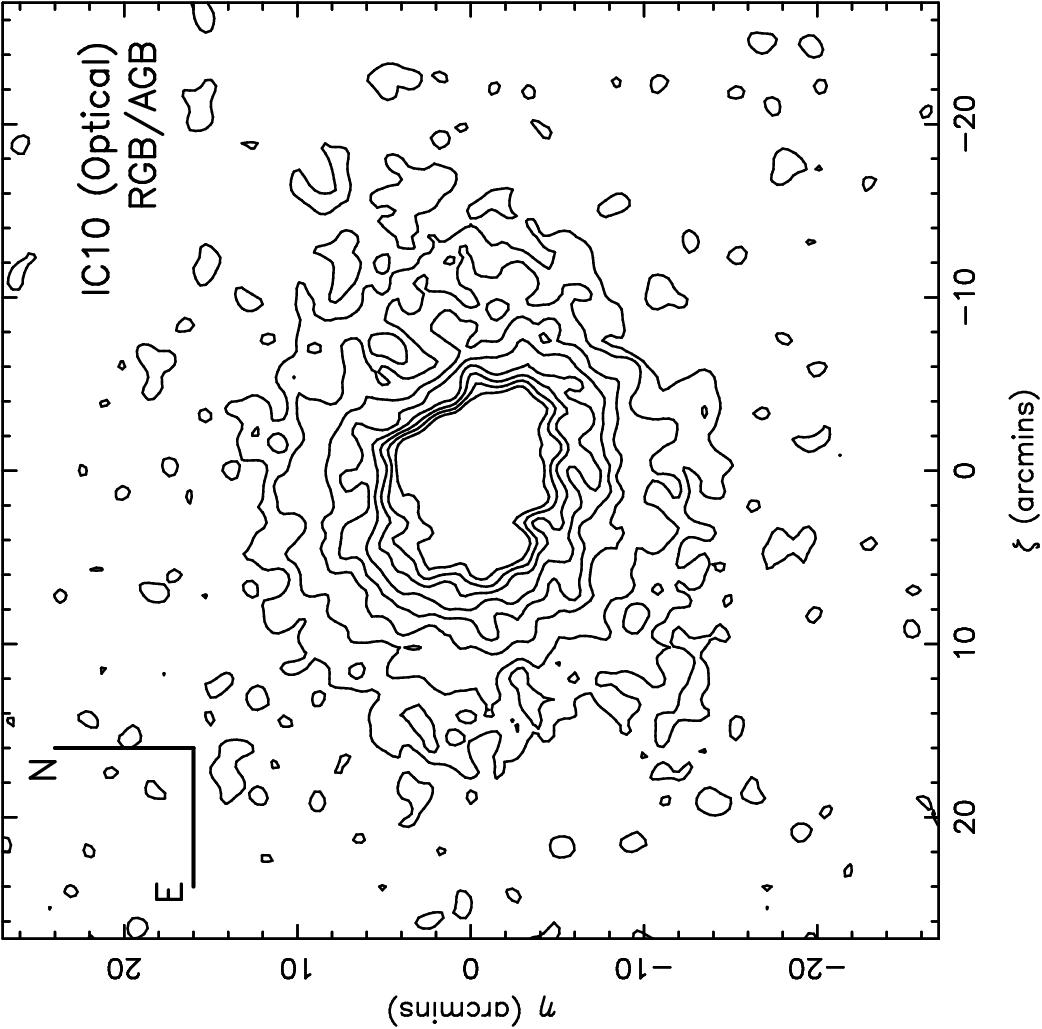}
   \caption {Left panel: blue points indicate the positions of candidate young stars in IC10, identified in the CFHT/MegaCam data using $(g-i)_0<0$ and $i_0\le 21$. Grey-scale shows candidate RGB/AGB stars, selected as described in the text. Pixels are $0.3\times0.3$\,arcmins, convolved with a Gaussian kernal of dispersion 1.5 pixels. The density map is shown with a square-root scaling. Right panel: contour map of the RGB distribution, where the density map is the same as in the right panel. Contours are set at $2-, 5-, 10-, 15-...35-\sigma$ above the background level. The center of IC10 is positioned at $\alpha=00h 20s 17.3m, \delta=59^\circ 18' 14"$. } 
  \end{center}
\end{figure*}

\begin{figure*}
  \begin{center}
    \includegraphics[angle=270, width=8.5cm]{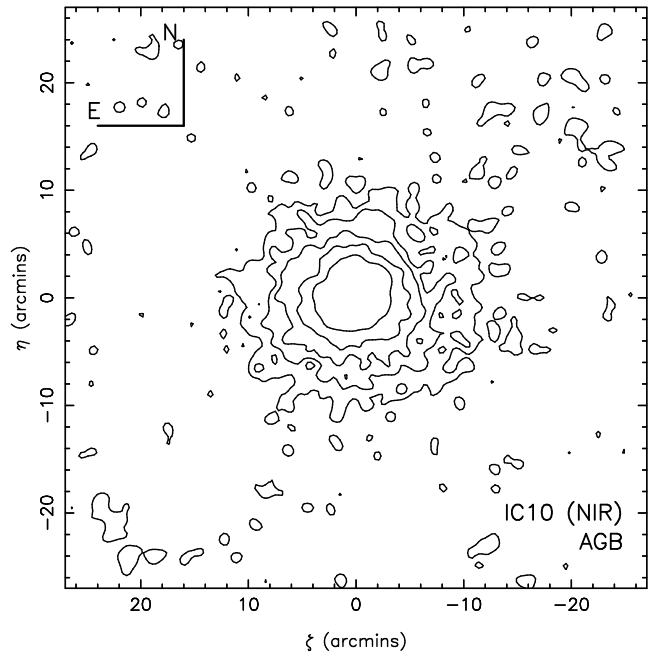}
   \caption {Contour map of the AGB distribution from the UKIRT/WFCAM data, selected as described in the text. Pixels are $0.3\times0.3$\,arcmins, convolved with a Gaussian kernal of dispersion 1.5 pixels. Contours are set at $2-, 5-, 10-, 20-, 40-, 60-, 80-\sigma$ above the background level. The center of IC10 is positioned at $\alpha=00h 20s 17.3m, \delta=59^\circ 18' 14"$.  } 
  \end{center}
\end{figure*}

\begin{figure*}
  \begin{center}
    \includegraphics[angle=270, width=8.5cm]{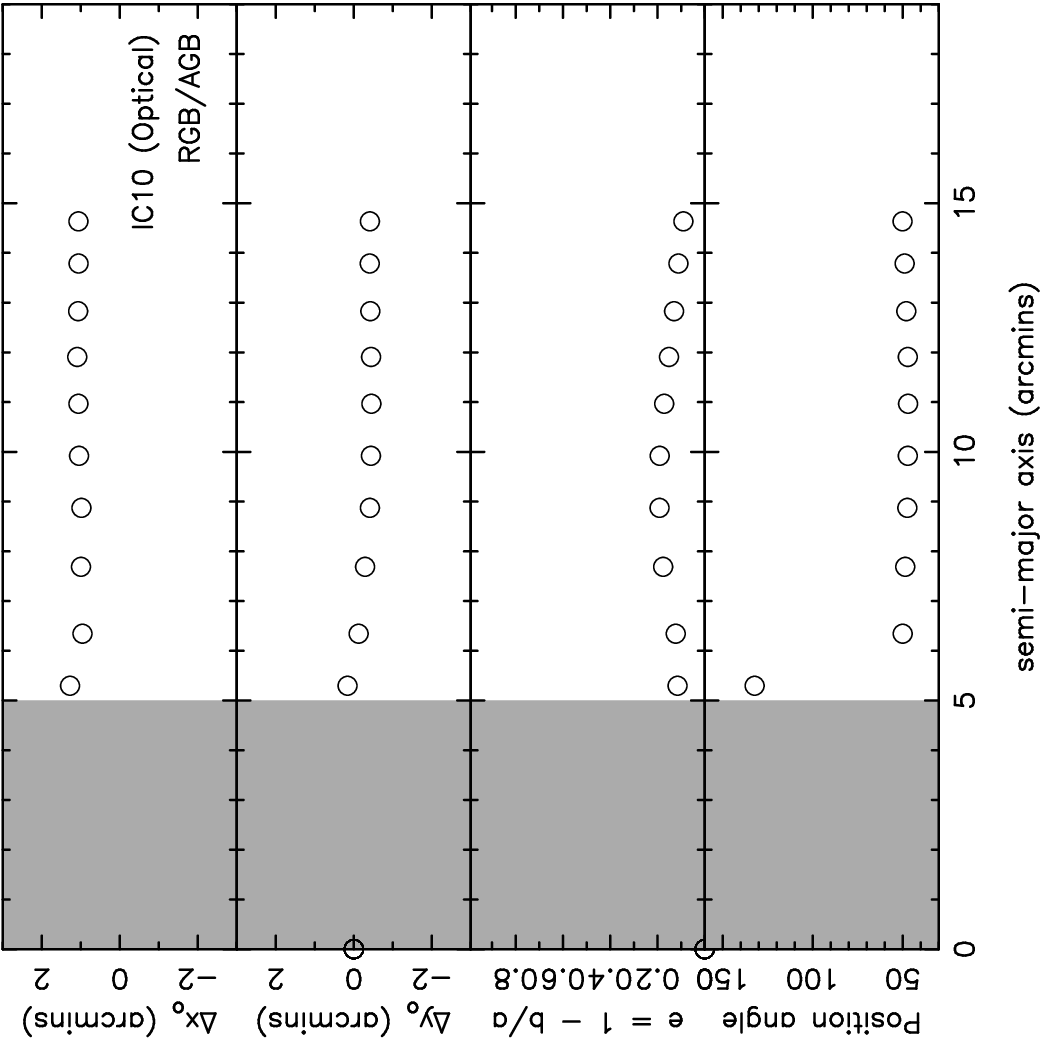}
    \includegraphics[angle=270, width=8.5cm]{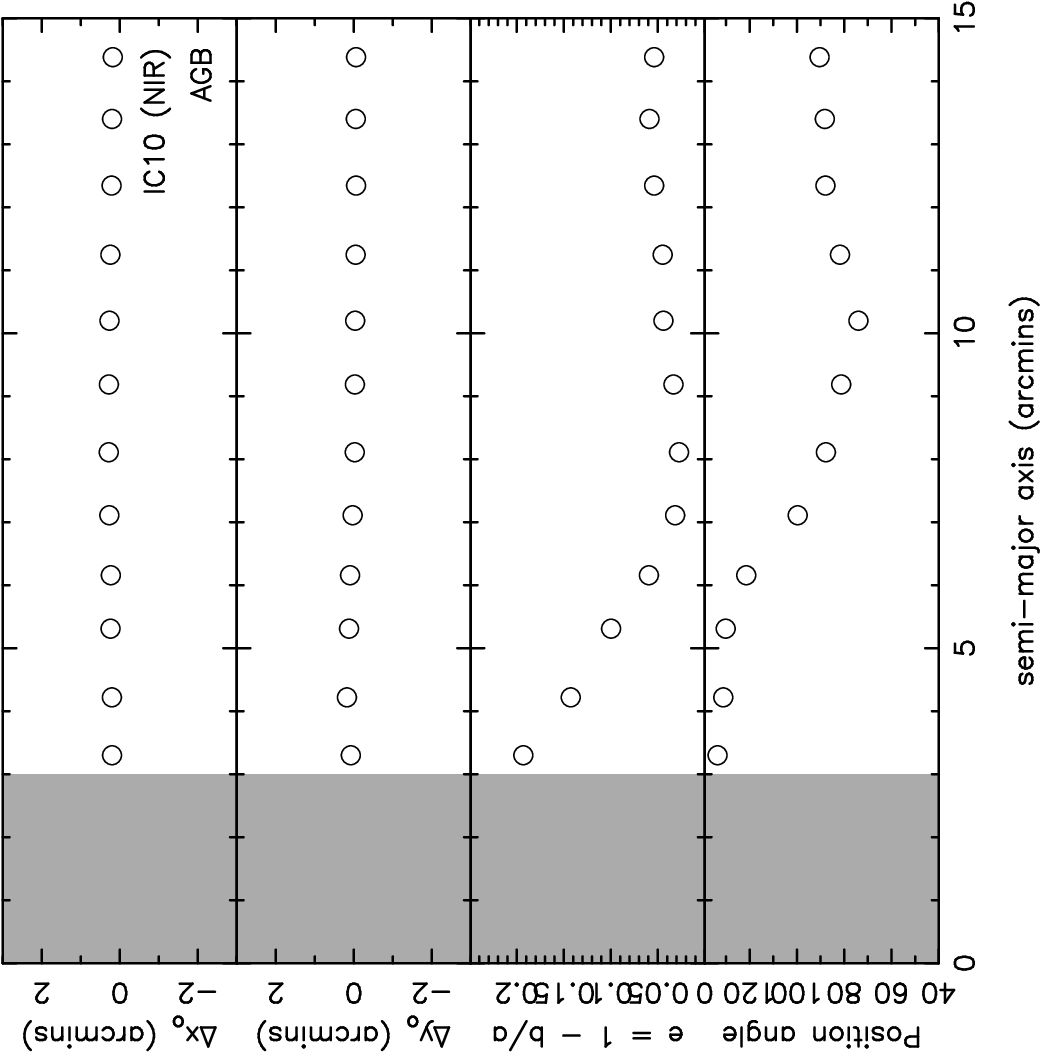}
   \caption {Variation in the shape of IC10 as a function of semin-major axis, as estimated from the RGB/AGB distribution using the CFHT/MegaCam data (left panels) and the AGB distribution using th UKIRT/WFCAM data (right panels). The variation in the center of the best fitting ellipse ($x_0, y_0$; top two panels), its ellipticity (third panel) and position angle (measured east from north; fourth panel) are plotted as a function of the semi-major axis. The grey region in the left hand plot indicates the region where estimates are less reliable due to crowding effects. See text for details.} 
  \end{center}
\end{figure*}

We now examine the spatial distribution of the three main stellar populations identified in our IC10 data, namely young blue stars, AGB stars (O-type and C-type), and RGB stars. AGB and RGB stars are present in both the optical and near-IR CMDs, but the bright AGB stars are most uniquely identified in the near-IR CMD. RGB stars, on the other hand, can be identified as being the majority of stars fainter than the TRGB in the near-IR CMD, but the depth of the WFCAM data means that this dataset is sub-optimal for examining the distribution of this population. In the optical data it is not possible to disentangle the AGB from the RGB due to the confusion introduced by the reddening, but weighed against this consideration is the fact that the optical data is considerably deeper, thus identifying considerable more RGB stars than in the near-IR.

The left panel of Figure~8 shows the spatial distribution of stars in the CFHT/MegaCam optical dataset. The grey-scale shows a density plot of candidate RGB/AGB stars, selected to have $(g-i)_0>0$ (to avoid young stars) and $i_0>20.4$ (to reduce foreground contamination). The large gaps that are present between the first-and-second, and third-and-fourth, rows of CCDs in MegaCam have been artificially filled in this image using estimates of the local background. Pixel size is 0.3\,arcmins, smoothed with a Gaussian kernal with a dispersion of 1.5 pixels. Blue points indicate candidate young stars, selected to have $(g-i)_0<0$ and $i_0\le 21$. The right panels show the same density map as shown in the grey-scale, but now as a contour map. The lowest level contour is set at $2-\sigma$ above the background, then $5-\sigma$, increasing to $35-\sigma$ in increments of $5-\sigma$. No contours are shown for the very central regions, since here there is extreme crowding and we focus our analysis on the outer regions only. Figure~9 shows the equivalent contour map for the UKIRT/WFCAM dataset, where only candidate AGB stars have been selected, using the foreground color-color cut described earlier and requiring $K_0<17.95$ (i.e., brighter than the expected magnitude of the TRGB). Pixel sizes and smoothing is the same as for the optical dataset. Contours are set at $2-, 5-, 10-, 20-, 40-, 60-, 80-\sigma$ above the background level.

We quantify the two-dimensional shape of the stellar population as traced in the contour maps of IC10 via a moments analysis of the density distribution. Specifically, we calculate the center of gravity $(x_o,\,y_o)$, position angle, $\theta$ (measured east from north), and eccentricity, $e$, via

\begin{eqnarray}
x_o&=&\frac{\sum_i x_i I\left(x_i,\,y_i\right)}{\sum_i I\left(x_i,\,y_i\right)}~~~~~y_o~=~\frac{\sum_i y_i I\left(x_i,\,y_i\right)}{\sum_i I\left(x_i,\,y_i\right)}\nonumber\\
~\nonumber\\
\theta&=&\frac{1}{2}\arctan\frac{2 \sigma_{xy}}{\sigma_{yy} - \sigma_{xx}}\nonumber\\
~\nonumber\\
e&=&\frac{\sqrt{\left(\sigma_{xx} - \sigma_{yy}\right)^2 + 4 \sigma^2_{xy}}}{\sigma_{xx} + \sigma_{yy}}
\end{eqnarray}

\noindent where $I\left(x_i,\,y_i\right)$ is the intensity of the
$i^{th}$ pixel and $\sigma_{xx}$, $\sigma_{yy}$ and $\sigma_{xy}$ are
the intensity-weighted second moments of the distribution. The
eccentricity is related to the ellipticity, $\epsilon = 1 - b/a$, via

\begin{equation}
\epsilon = 1 - \sqrt{\frac{1 - e}{1 + e}}~.
\end{equation}

\noindent These quantities are calculated as a function of radius using an iterative technique. That is, for all pixels lying within a given radius $r$ of a first estimate of the center, we calculate the corresponding value of $x_0, y_0, e, \theta$. We then repeat the calculation but this time only using pixels lying within the appropriately positioned ellipse described by these parameters (where $r$ now describes the semi-major axis of the ellipse). This process is repeated until convergence for each value of $r$. Results are shown in Figure~10 for the optical data (left panel) and the near-IR data (right panel). For each stellar population, the estimates within the inner few arcminutes are either unreliable or impossible due to the effects of crowding; these regions are marked by the shaded areas in Figure~10. In these plots, $\Delta x_0, \Delta y_0$ corresponds to the offset from an adopted center for IC10 of $(00^h20^m17.3^s, 59^\circ18'14")$. 

Table~1 lists the average values for the various parameters for each stellar population. This table also includes an estimate of the position of the center of the young population. Due to the limited number of young stars, the center of the population is estimated as the median position of all stars within 10 arcmins of the nominal (literature) center of IC10. We find that the center of the young stars is within arcseconds of the nominal center, which is entirely consistent with the expectation that the  recognised ``center'' of IC10 coincides with the center of the stellar population that completely dominates the light. The scale radius has been crudely estimated as the median distance of these stars from the center of the population.

\begin{table*}
\begin{center}
\begin{tabular*}{0.8\textwidth}{r|ccccccc}
&\multicolumn{2}{c}{$(\Delta x, \Delta y)$}& $\theta$ &  $\epsilon$ & n & \multicolumn{2}{c}{$r_e$}\\
& arcsecs & pc ($D/750$kpc) & $\circ$ & $1 - b/a$ &  &arcmins &  pc ($D/750$kpc)\\
\hline
RGB/AGB (optical) & (62.1, -22.7) & (226, -83) & 48 & 0.15 & $0.75 \pm 0.06$ &$5.75\pm 0.11$ & $1254 \pm 24$ \\
AGB (near IR) & (13.6, 0.7) & (49, 3) & 100 & 0.07 & $1.20 \pm 0.17$ &$3.22\pm 0.30$ & $ 702 \pm 65$\\
Young (optical) & (1.6, 1.3) & (6, 5)&---& ---& --- & $\sim 1.75$ & $\sim 382$ \\
\hline
\end{tabular*}
\caption{Shape and structural parameters for each of the main stellar populations identified in IC10. Positional offsets are measured relative to the literature center for IC10 of $(00^h20^m17.3^s, 59^\circ18'14")$, derived  from the 2MASS Extended Source catalog. }
\end{center}
\end{table*}

\subsection{Radial profiles}

\begin{figure*}
\begin{center}
\includegraphics[angle=270, width=8.5cm]{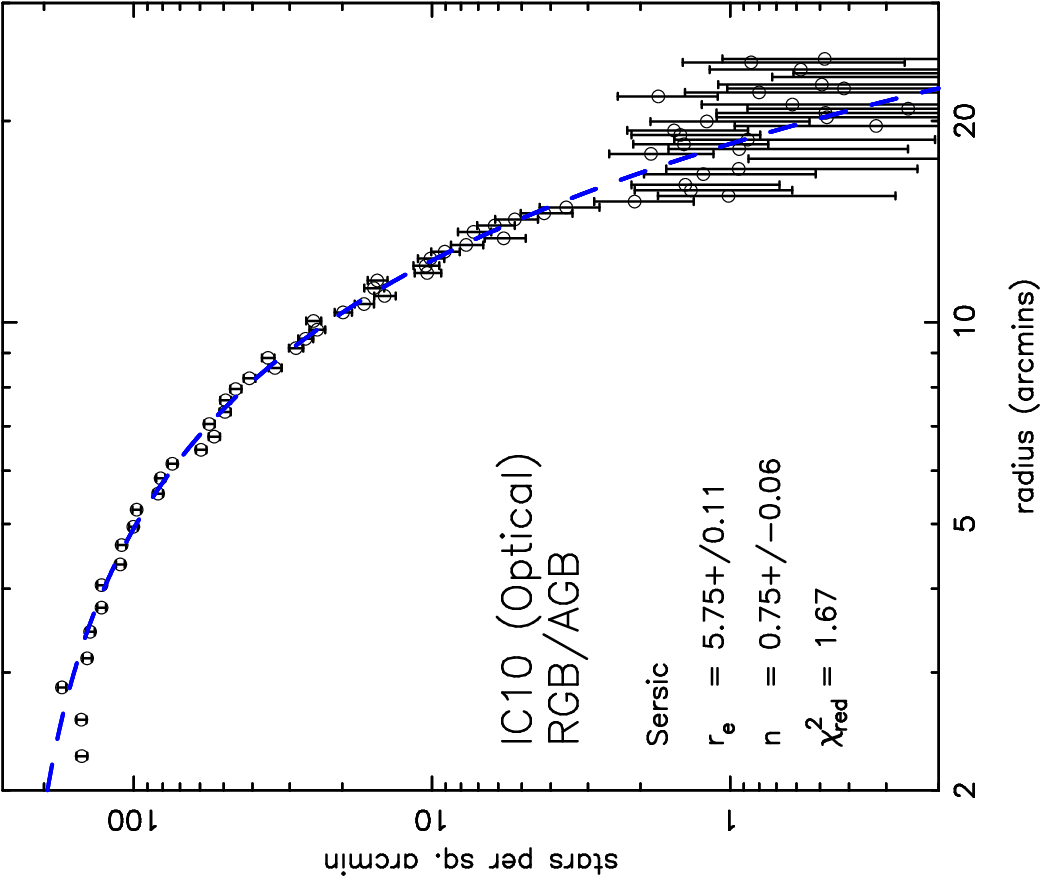}
\includegraphics[angle=270, width=8.5cm]{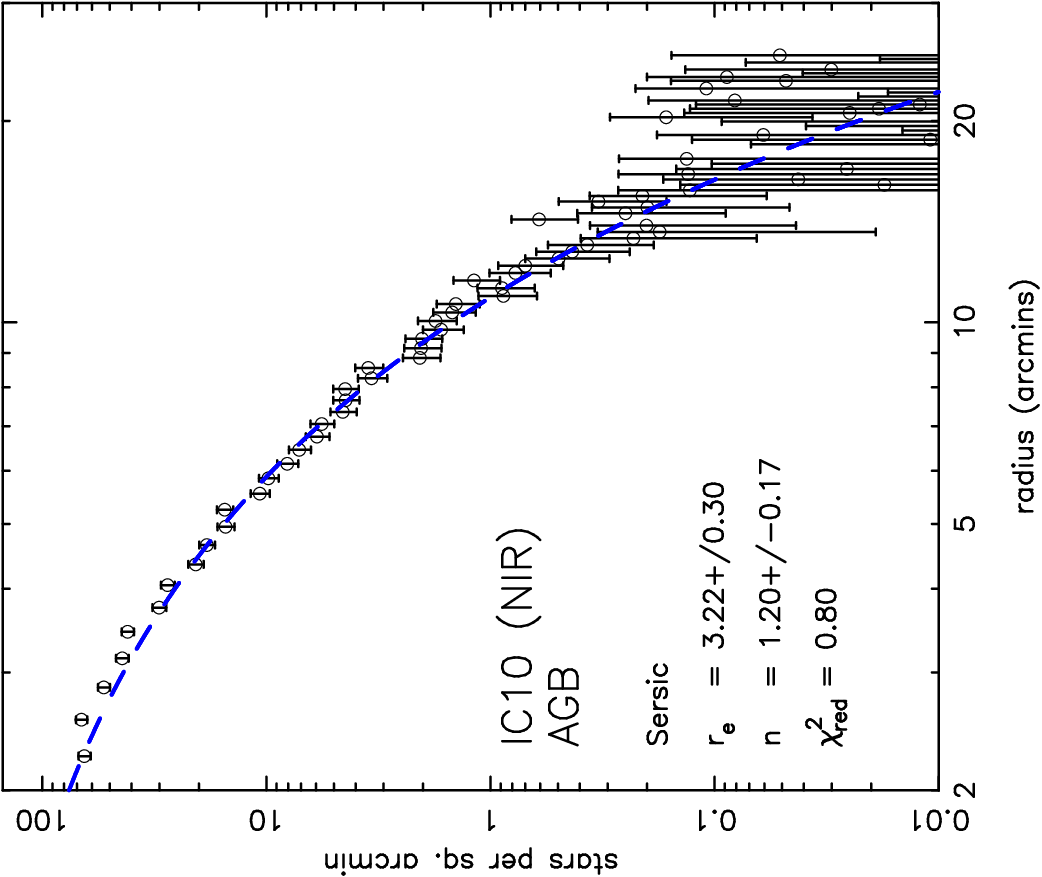}
\caption {Radial profiles of the RGB/AGB population from the CFHT/MegaCam data (left panel) and of the AGB population from the UKIRT/WFCAM data (right panel), along with best-fitting Sersic profiles (dashed lines). For both panels, the central regions where crowding is significant have been omitted from the fits. } 
\end{center}
\end{figure*}

We now calculate the radial profiles of the RGB/AGB population shown in Figure~8, and the AGB population shown in Figure~9. As shown in Figure~10 and Table~1, both populations are relatively circular. To facilitate comparison between the two profiles, we therefore adopt circular annuli centered on the nominal center of IC10 from the literature $(00^h20^m17.3^s, 59^\circ18'14")$.  Background levels are estimated in the outer parts of the fields, and subtracted from the profiles. We also apply a crowding correction, following \cite{irwin1984}:

\begin{equation}
f^\prime \simeq f \left( 1 + 2 f A^\prime +\frac{16}{3} f^2 A^{\prime 2} ... \right)~.
\end{equation}

\noindent $f$ is the observed number density of all images, $f^\prime$ is the actual number density and $A^\prime$ is the typical area of the image, where the radius of the image is closely approximated by the seeing. This correction ignores second-order effects such as the actual shape of the luminosity function. Clearly, this correction is only applicable in partially crowded fields, and breaks down once crowding becomes sufficiently severe.

Radial profiles for IC10 are shown in Figure~11. The left panel corresponds to the RGB/AGB population (optical), and the right panel corresponds to the AGB (near-IR) population. Error bars indicate the combined uncertainty of the counts and the background estimate. Only the data for regions beyond 2 arcminutes are shown, since in both datasets, crowding is severe interior to this radius. Both profiles are remarkably regular, with no obvious breaks or discontinuities. 

We fit a Sersic function to each radial profile, given by

\begin{equation}
I(r) = I_0 \exp\left[-b_n\left(\frac{r}{r_e}\right)^{\frac{1}{n}}\right]~.
\end{equation}

\noindent $I_0$ is the central intensity, $r_e$ is the effective radius, $n$ is the Sersic index, and $b_n$ is a quantity that depends upon $n$ and is well approximated by $b_n\simeq 1.9992n-0.3271$ for $0.5\le n \le 10$ (e.g., \citealt{graham2005}). Best fit values for each profile are included in each panel, and summarised in Table~1.

\section{Discussion}

\subsection{The distribution of stars through time}

Over the past decade, it has become increasingly clear that dwarf galaxies exhibit complex spatial structures in terms of age, metallicity and chemical variations. For example, \cite{harbeck2001} showed that population gradients are common in the Local Group dwarf galaxy population. \cite{tolstoy2004} subsequently showed that the prototypical Sculptor dwarf spheroidal galaxy consists of two dynamically and chemically distinct components, wherein the younger, more metal-rich population is more centrally concentrated than the older, more metal-poor, component. It is now generally accepted that different stellar populations within the same dwarf galaxy can have distinct spatial distributions (e.g., \citealt{battaglia2006, battaglia2011, battaglia2012, bernard2008, lee2009}, among many others).

Based on our analysis, we confirm that IC10 exhibits the behaviour of stellar populations seen in many other Local Group gaaxies. Firstly, there is a clear age gradient, such that the young main sequence and blue loop stars are the most concentrated of the stellar populations identified in our data, and likely have ages in the range of tens to a few hundred Myrs (see also \citealt{massey2007}). Secondly, bright AGB stars, that may have ages as young as $\sim1$\,Gyr, have an effective radius that is of order twice as extended as the youngest stars. Finally, the RGB stars, that are generally old ($>> 2 - 3$\,Gyrs), have an effective radius that is roughly 4 times as extended as the youngest population. While typical of Local Group galaxies, this behaviour is also typical of other systems classified as blue compact galaxies, that are known to be embedded within extended low surface brightness distributions of old stars (\citealt{loose1986}).

While each stellar population has a different characteristic scale radius, the center of the young population in IC10 is clearly offset from the oldest RGB population. Unsurprisingly, the centroid of the young stars is very close to the literature value for the central coordinates of IC10, since the latter are based on the centroid of the light distribution (measured from the 2MASS All-Sky Release Point and Extended Source Catalogs and Image Atlas\footnote{http://www.ipac.caltech.edu/2mass/releases/allsky/index\_extmis.html}), which is dominated by the contribution of the starburst population. The centroid of the AGB population is also close to this position. However, the center of the oldest stellar population - the RGB - is very significantly different from these coordinates. As listed in Table 1, the offset of the RGB centroid to that of the young stars is $\sim 240$\,pc, for an adopted IC10 distance of 750kpc. To put the significance of this number in context, the half-light radius of the Sculptor dwarf spheroidal galaxy is $\sim 280$\,pc.

As the oldest stellar population, RGB stars are expected to be the most relaxed of the stellar populations that we examine. As such, their centroid is likely close to the centroid of the net gravitational potential of the galaxy. Star formation in dwarf galaxies need not be a uniform and symmetric process. Indeed, an irregular distribution of young stars is a defining morphological characteristic of Magellanic irregular galaxies (\citealt{holmberg1958}), and IC10 was first classified as such a system by \cite{devaucouleurs1965}. \cite{bastian2011} show, through tracing blue loop stars, that star formation in dwarf galaxies is highly substructured on scales of tens to hundreds of parsecs, and that such substructures can persist for potentially hundreds of megayears (see also \citealt{dohmpalmer2002}). In this picture, it is not surprising that the nearest galaxy with the highest star formation rate has ongoing star formation that is not centered in the gravitational potential of the galaxy. Indeed, if the star formation in IC10 was induced through an interaction, then it is certainly not clear that the star formation will be centered with respect to the net gravitational potential. 

Finally, we suggest that it is important that the difference between the center of the young stellar population in IC10 and that of the old population is considered when conducting dynamical analyses of IC10 in which a model of the gravitational potential is being used. Specifically, large systematic uncertainties could be introduced if models fail to properly account for the fact that the center of mass of this galaxy is not the same as its center of light.

\subsection{The outer structure and interaction history of IC10}

\cite{loose1986} demonstrated that blue compact galaxies consist of at least two components; the central, star-bursting population and an old, extended, low surface brightness surrounding population. \cite{sanna2010} showed that red giants surround IC10 extending to large radius, and here we measure the shape parameters describing this stellar component. In addition, we also show that stars with ages intermediate between the youngest and oldest populations have a spatial extent that is also intermediate between these populations.

The spatial extent of the outer component of RGB stars in IC10 is worthy of comment. Firstly, the best-fit Sersic profile to the RGB distribution yields $n= 0.75 \pm 0.06$. In this respect, this component of IC10 is typical of many dwarf galaxies for which the stellar distribution falls off more slowly than an exponential (e.g., \citealt{ferrarese2006b}). With an effective radius of $r_e = 5.75\pm 0.11$\,arcmins ($\sim1.2$\,kpc for an assumed IC10 distance of 750kpc), very few dwarf galaxies in the Local Group are as large as this old component of IC10. Comparison with the updated data tables presented in \cite{mcconnachie2012}\footnote{Available at http://www.astro.uvic.ca/$\sim$alan/Nearby\_Dwarfs\_Database.html} show that (excluding the LMC, SMC and M33) there are only 6 Local Group galaxies with a robust measurement of their size that places them with $r_h > 1$kpc (measured on the major axis). Of these, the Sagittarius dwarf spheroidal is known to be exceptionally tidally perturbed (\citealt{ibata2001b, majewski2003} and many others). Four others in this sample are M31 satellites (Andromeda II, XIX, XXIII and XXXII), and each of these is also a dwarf spheroidal. IC10 is also a satellite of M31, but is not classed as a dwarf spheroidal due to its young stars and gas. The remaining galaxy in this sample is WLM, a relatively isolated dwarf irregular galaxy of comparable luminosity to IC10. 

It is important to note that, in addition to a large spatial extent, the RGB population (and also the AGB population traced with the near IR data) has a very regular morphology. This is visible in both panels of Figure~8, where the isophotes do not show any obvious signs of distortion, or any features that might suggest the presence of stellar streams or shells. Figure 10 shows the ``intensity-weighted'' shape  of the galaxy as a function of radius as traced by the stellar populations. At large radius, IC10 is well approximated as circular with no evidence of significant isophote twisting or other irregular behavior. There is similarly a lack of any notable features at large radius visible in the radial profiles in Figure~11. In both the optical and near-IR data, the decline in star counts matches well the Sersic profile fits. There is no sign that the profiles truncate at any point, and so it is probable that stars will be found at even larger radius than probed by our data. There is also no clear sign of any ``extra-tidal stars'', that is, stars located at large radius in numbers exceeding that expected from a simple extrapolation from the numbers at smaller radii. The presence of such stars has been suggested in the past to be due to tidal effects (\citealt{irwin1995}), although other explanations such as multiple components are also possible. We note that there is a slight deviation of the RGB star counts from the fitted profile in the left panel of Figure~11 at around 15 arcmins; this could plausibly be interpreted as marking the truncation of IC10 with the stars at larger radii being ``extra-tidal''. However, such an interpretation is at the limits of the data, and we consider it unlikely given that no similar feature is observed in the near IR data. We conclude that, given the data in hand, the outskirts of IC10 are unremarkable in their regularity, which is itself of interest in comparison to the HI distribution. 

\cite{demers2004} analyse the structure of IC10 using CFHT/CFHT12K wide field photometry ($42' \times 28'$) in the Mould $I$ and $R$ filters and narrow-band $CN$ and $TiO$ filters, that they use to investigate the spatial distribution of Carbon stars and the rest of the red giant population. They conclude that the C stars follow the bulk of the rest of the stellar population of IC10, and that both are well described by power law profiles. Our Sersic profiles in the radial range of ~5 - 14 arcmins from the center of IC10 (a region in common with \cite{demers2004}) are broadly consistent with the power law distributions discussed in \cite{demers2004}. However, we also trace these populations to even larger radius due to the larger field of view of MegaCam compared to CFHT12K, implying an even larger/more extended system as discussed above. The larger radial baseline allows us to show that the AGB population is more concentrated overall than the RGB, in contrast to the conclusion of \cite{demers2004}. These authors also suggest that the inner distribution of red giants is very irregular, with a ``horseshoe-like'' shape. We are unable to comment on the very central distribution of red giants in our data, since the natural seeing of the site and the very bright starburst population in this region means we have severe crowding/incompleteness that affects any analysis we would conduct in this region.

\cite{nidever2013} present strong evidence for an extended HI feature stretching 1.3\,degrees to the northwest of IC10. This feature  displays a large velocity gradient that ends in a high column-density clump, and \cite{nidever2013} argue that it is plausibly the result of tidal interactions in a dwarf-dwarf merger. If this proposed merger is also the cause of the strong star formation in IC10, then the merger is relatively recent, and it is reasonable to suppose that the merger could leave an imprint on the stellar distribution. However, no such features are observed in the optical and NIR datasets analysed here, and there are no obvious correlations with the morphology of the gas. For example, the stellar distribution is not obviously aligned with the direction of the extended HI, nor are there obvious extensions in the stellar contours that could be a sign of a recent interaction (either through distortion of the main body of IC10 or through stars left by a tidal stream from the companion). However, we also stress that the current optical and NIR data do not cover a large enough spatial extent to match with the position of the proposed progenitor (the optical data extends $\sim0.5$\,degrees away from the center of IC10, whereas the HI extends for 1.3 degrees). It is clearly worthwhile to obtain both deeper and wider data to fully explore the interaction scenario and even potentially uncover a new companion to IC10. Our current data covers a similar radial extent of earlier HI studies, for example by \cite{wilcots1998}, and we refer the reader to their Figure 2 for a column density map of the HI in IC10. Again, no clear correlations are present between the structure in the HI and the (lack of) structure in the older stellar distributions.

\section{Summary}

In this paper, we have used optical and NIR wide field imaging data of the nearby, Local Group star-bursting, blue compact dwarf galaxy IC10 to investigate the outer stellar structure and older stellar populations of this unusual and highly obscured galaxy. Using CFHT/MegaCam and UKIRT/WFCAM we have examined the spatial distribution of the various stellar populations present, tracing them to large radius and low surface brightness. A large number of bright AGB stars prevent us from obtaining a precise distance to this galaxy, but we constrain it to lie in the range of 700 -- 800kpc, consistent with earlier studies. Stellar populations of decreasing average age are increasingly spatially concentrated in IC10, in line with results for many other Local Group galaxies and consistent with expectations for blue compact galaxies that have the star-forming region embedded within a much larger, lower surface brightness component of older stars. The youngest populations (those of the star burst and the AGB stars) are significantly offset from the geometric center of the older RGB population, implying that recent star formation occurred significantly away from the center of the galaxy. The scale of this offset is significant such that dynamical models of the evolution of IC10 should distinguish between the center of light of this galaxy and its center of mass. The RGB population falls off slower than an exponential and has an effective radius of approximately 5.75\,arcmins ($\sim1.2$\,kpc), making this blue compact galaxy one of the most extended of the known dwarf galaxies in the Local Group. Importantly, the outermost isophotes of this galaxy are very regular in shape and essentially circular in morphology, and we find no evidence in the extended stellar component to suggest that IC10 has undergone a recent interaction as proposed by \cite{nidever2013}. However, the footprints of the optical and near-IR data do not cover as large an area as the HI data, and it is possible that stellar signatures of this proposed encounter exist outside of the survey region, demonstrating the need for deeper and wider surveys of IC10.

\section*{Acknowledgments}

We thank Kristy McQuinn and Evan Skillman for useful discussions, including highlighting relevent references, and for reading an early draft of the manuscript. We thank the referee for very useful comments that helped improve this paper.

\end{document}